\documentstyle[12pt]{article}

\def\prd#1{{\em Phys.~Rev.}~{\bf D#1}\ }
\def\prl#1{{\em Phys.~Rev.~Lett.}~{\bf #1}\ }
\def\plett#1{{\em Phys.~Lett.}~{\bf #1B}\ }
\def\np#1{{\em Nucl.~Phys.}~{\bf B#1}\ }
\def\deg{\ifmmode{^{\circ}}\else ${^{\circ}}$\fi}

\def\bi{\begin{itemize}}
\def\ei{\end{itemize}}
\def\ed{\end{document}}
\def\be{\begin{equation}}
\def\ee{\end{equation}}
\def\beq{\begin{eqnarray}}
\def\eeq{\end{eqnarray}}

\def\bm#1{\ifmmode{\mbox{\boldmath $#1$}}\else {\boldmath $#1$}\fi}

\def\eb{\end{thebibliography}}

\def\mueg{\ifmmode{\mu\rightarrow e\gamma} \else $\mu\rightarrow e\gamma$\fi}
\def\taumg{\ifmmode{\tau\rightarrow \mu\gamma}\else
$\tau\rightarrow \mu\gamma$\fi}
\def\taueg{\ifmmode {\tau\rightarrow e\gamma}\else
$\tau\rightarrow e\gamma$\fi}

\begin{document}

\begin{titlepage}
\begin{flushright}

hep-ph/9801348{\hskip.5cm}\\
\end{flushright}
\begin{centering}
\vspace{.3in}
{\bf A NOTE ON $R$-PARITY VIOLATION AND FERMION MASSES }\\
\vspace{2 cm}
{M.E. G\'OMEZ and K. TAMVAKIS}\\
\vskip 0.5cm
{\it Division of Theoretical Physics,}\\
{\it University of Ioannina, GR-45110, Greece}\\ \vspace{0.5cm}

\vspace{1.5cm}
{\bf Abstract}\\
\end{centering}
We consider a class of supersymmetric 
$SU(3)\times SU(2)\times U(1)$ 
multihiggs models in which $R$-parity is violated 
through bilinear Higgs-lepton interactions. 
The required, due to $R$-parity violation, higgs-lepton rotations introduce 
an alternative way 
to generate the phenomenologically desirable fermion mass matrix 
structures independently of the equality of Yukawas, 
possibly imposed by superstring or other unification.

\vspace{.1in}

\vspace{1cm}
\begin{flushleft}

January 1998\\
\end{flushleft}
\hrule width 6.7cm \vskip.1mm{\small \small}
E-mails\,:\, mgomez@cc.uoi.gr, tamvakis@cc.uoi.gr

\end{titlepage}
One of the many consequences of modern unified theories are the 
relations they imply 
among Yukawa couplings. Such relations exist in GUT models\cite{XYZ} 
as well as in models derived in the context of 
Superstring Theory\cite{XYW}. These relations reduce the number of free parameters 
required to fit the fermion masses and mixing angles in the framework 
of the Standard Model. The theoretical and phenomenological success of 
unification ideas in explaining certain parameters of the Standard Model 
has made the {\it{supersymmetric unification}}\cite{XYH}\cite{XYG} framework very attractive despite 
the fact that some of the predicted mass relations are not automatically 
successful in a minimal context. In SUSY models
certain Ans{\"{a}}tze incorporating ``texture'' zeroes at superlarge
 energies have 
been proven to be an effective method to explain fermion masses 
while reducing the number of free parameters \cite{FR}.

Simple $SU(5)$ requires at the unification scale
 $\lambda_{b}=\lambda_{\tau}$. This leads to the experimentally observed
 $m_b/m_{\tau}$ 
mass ratio \cite{CH}. Nevertheless, the identification of the down  and lepton Yukawa 
couplings at high energies does not lead to the correct prediction for 
the lightest generation 
mass ratios. A solution to this problem was proposed initially by Georgi and
Jarlskog \cite{GJ}. Ramond, Roberts and Ross \cite{RA} performed a systematic 
search for
symmetric mass matrices with a maximum number of texture zeroes and were led to 
several satisfactory Ans{\"{a}}tze. In these proposals the fermion mass ratios 
are brought in agreement with the experimental data by taking the lepton Yukawa 
matrix to be almost identical to the down Yukawa matrix apart from a factor of 
three in (2,2) entry. Several attempts have been made to incorporate such 
Ans{\"{a}}tze and explain the down/lepton 
relative factor of three in $SO(10)$ and $SU(5)$ models 
introducing extra Higgses 
in various representations \cite{AN}. Here we present an alternative approach 
in which the relative factor of three 
in the lepton matrix is generated as a consequence 
of field redefinitions required by the presence of interactions
 that violate  $R$-parity.

In the five solutions presented in ref.~\cite{RA} the down quark Yukawa matrix takes
the following form
\be
Y_d =  \left( \begin{array}{lll} 0 & F & 0\\ 
F^{\ast} & E & E^{\prime}\\
0 & E^{\prime} & D \end{array} \right) 
\ee 
Where $E^{\prime}$ can be taken either as $0$ or of the same order 
of magnitude as $E$ , depending on the choice of up-quark Yukawa matrix. 
The lepton Yukawa matrix is taken to be
\be 
Y_e =  \left( \begin{array}{lll} 0 & F & 0\\ 
F^{\ast} & 3 E & E^{\prime}\\
0 & E^{\prime} & D \end{array} \right) 
\ee
The elements of the above matrix obey the approximate relations
\be
\frac{F}{E}=\lambda\,\,\,\,\, \ \ \frac{E}{D}= 2\lambda^3\,\,\,\,\,\ \ \ 
 E\approx E^{\prime} 
\ee
 $\lambda \approx .22$ stands for the {\it{Cabbibo angle}}.

Since $D >> E, E^{\prime} $ , the diagonalization 
of the matrix corresponding to 
the first two generations gives a ratio of masses 
in good agreement with the experimental data
\be 
\frac{m_d}{m_s}=9 \frac{m_e}{m_\mu}
\ee
Note that for this kind of textures proposed in ref.~\cite{RA} the masses of 
the first and 
second generations are independent of the third generation Yukawa. 
In what follows 
we shall consider a two generation model
  in which down quark and lepton masses are described by  matrices
\be
{\cal{D}}= \left( \begin{array}{ll} 0 & F\\ F & E\end{array} \right)
\,\,\,\,\,\,  
{\cal{E}}=\left(  \begin{array}{ll} 0 & F\\ F & -3E\end{array} \right)
\ee

All the above parameters are taken to be real.

Let us now consider a two generation model with two pairs of Higgs isodoublets. 
The down quark and lepton Yukawas are equal as is the case in a large class of 
unified models.
 $R$-parity is broken in this model by bilinear terms in the superpotential
\be
W= \mu_i h_i h_i^c + \xi_{ij}h_i^c l_j + Y_{ijk} \left( q_i d_j^c + l_i e^c_j
\right) h_k
\ee
Where $i, j, k =1,2 $. We shall take the $\xi$ -matrix to be
\be
\xi_{ij}=\left( \begin{array}{ll} 0 & \xi_2 \\
                                 \xi_1 & 0 \end{array} \right)
\ee

At this point let us introduce the angles
\be
\sin \theta_1 =\frac{\xi_2}{\sqrt{\mu_1^2 + \xi_2^2}} \ \ \ 
\sin \theta_2 =\frac{\xi_1}{\sqrt{\mu_2^2 + \xi_1^2}}
\ee
and the {\it{Higgs mass-eigenstate fields}}
\beq
 H_1 = \cos\theta_1 h_1 + \sin\theta_1 l_2 \\
 H_2= \cos\theta_2 h_2 +\sin\theta_2 l_1\\
 L_1 = -\sin\theta_1 h_1 + \cos\theta_1 l_2 \\
 L_2=-\sin\theta_2 h_2 + \cos\theta_2 l_1
\eeq

 Since our model serves only a demonstrative purpose it is not restrictive to 
 assume that the parameters $\mu_i$ and $\xi_i$ are of the same order of magnitude 
 $M_W$ and 
chose angles $\theta_1 \approx \theta_2 = \theta $. 
Note that for $\xi_1=\xi_2=\mu_1=\mu_2$ , $\theta=\pi/4$. 
The superpotential becomes
\beq
W &=& \sqrt{\mu_1^2 + \xi_2^2} H_1 h_1^c +  \sqrt{\mu_2^2 + \xi_1^2} H_2 h_2^c 
+  Y_{ijk}\cos\theta H_k q_i d_j^c 
\nonumber \\
&+& \left[ Y_{2j1}\cos{2 \theta} H_1+ \left(  Y_{2j2}\cos^2\theta - 
Y_{1j1}\sin^2\theta \right) H_2 \right] L_1 e_j^c \\ \nonumber
 &+&\left[\left( Y_{1j1}\cos^2\theta - Y_{2j2}\sin^2\theta\right) H_1 + 
Y_{1j2}\cos{2 \theta}  H_2 \right] L_2 e_j^c + W_{\Delta R}\\ \nonumber
\eeq
$W_{\Delta R}$ includes $R$-parity violating terms:
\be
W_{\Delta R}= -sin\theta Y_{ijk} q_i d^c_j L_k + 
sin\theta cos\theta (Y_{1j1}+Y_{2j2})(H_1H_2e^c_j -L_1L_2e^c_j)
\ee 
 $R$-parity violating terms of the type $qd^cL$ ,
  although phenomenologically chalenging, are not dangerous in
   themselves for proton decay, provided no bare terms of the type 
 $u^cd^cd^c$ exist. In an $SU(5)$ version of the Lagrangian (6)
  terms like $\xi_{ij}D_{Hi}d^c_j +Y_{ijk}(u_i^cd_j^c\overline{D}_{Hk}+
  q_il_j\overline{D}_{Hk})$ should be present arising from 
${\overline{\bf 5}}\ {\bf 5}_H+{\bf 10}\ {\overline{\bf 5}}\ {\overline{\bf 5}}_H$, 
$D_H$ and $\overline{D}_{H}$ denote the color triplets contained in ${\bf 5}_H$
and $\overline{\bf 5}_H$ respectively   . Models of this kind have been analyzed in \cite{KT}. 
  In contrast, 
 $H_1H_2 e^c_j$ terms are dangerous and should not be present. Note however 
 that due to $SU(2)$ antisymmetry these terms 
 are not present in a one-Higgs model.
Considering the vev's 
\beq
\left< H_1 \right> = v_1 \,\,\,\,\,\left< H_2 \right> = v_2 
\eeq

and defining
\beq
v_1 \ Y_{ij1} = Y_{ij}\,\,\,\,\,\,v_2 \ Y_{ij2} = I_{ij} \,\,\,\,\,\,
\alpha=\frac{v_2}{v_1}
\eeq
we see that these terms are eliminated if the following conditions are met
\beq
\label{eq:dt}
\alpha Y_{11} + I_{21} &=& 0 \nonumber \\ 
\alpha Y_{12} + I_{22} &=& 0 
\eeq
In the case $\xi = 0$, or equivalently for $\sin \theta =0$ and $L_1=l_2, L_2 = l_1$ 
the down quark and lepton masses are given by
\be 
M^d = M^e=Y_{ij} + I_{ij}
\ee

For non-zero $\xi$ these matrices are modified to
\be
M_{ij}^d = \cos{\theta} \left( Y_{ij} + I_{ij} \right)
\ee
and
\be
M^e = \left(\begin{array}{ll}
\cos2 \theta Y_{21} -\sin^2\theta \alpha Y_{11} + \cos^2\theta I_{21} &
\cos2 \theta Y_{11} -\sin^2\theta \alpha Y_{12} + \cos^2\theta I_{22} \\
\cos^2\theta Y_{11} -\frac{\sin^2\theta}{\alpha} I_{21} + \cos2\theta I_{11}& 
\cos^2\theta Y_{12} -\frac{\sin^2\theta}{\alpha} I_{22} + \cos2\theta I_{12}
\end{array} \right)
\ee

Remember that the desired structure of the fermion mass matrices is
\be{\cal{D}}= \left( \begin{array}{ll} 0 & F\\ F & E\end{array} \right)
\,\,\,\,\,\,  
{\cal{E}}=\left(  \begin{array}{ll} 0 & F\\ F & -3E\end{array} \right)
\ee

In order to match that by the derived mass matrices we must have 
four equations from the down quark mass
\be
M_{ij}^d={\cal{D}}_{ij}= \cos\theta (Y_{ij}+I_{ij})
\ee 

plus four additional equations derived from the lepton mass matrix
\beq
\label{eq:meq}
\cos2 \theta Y_{21} -\alpha \sin^2\theta Y_{11} + \cos^2\theta I_{21} &=& 0\\
\cos2 \theta Y_{22} -\alpha \sin^2\theta Y_{12} + \cos^2\theta I_{22} &=& F\\
\cos^2\theta Y_{11} -\frac{\sin^2\theta}{\alpha} I_{21} + \cos2\theta I_{11}
&=& F\\
\cos^2\theta Y_{12} -\frac{\sin^2\theta}{\alpha} I_{22} + \cos2\theta I_{12}
&=& -3E.
\eeq

Note that
\be
\label{eq:ccd1}
I_{ij}= \frac{{\cal{D}}_{ij}}{\cos\theta} -Y_{ij}
\ee
After inserting this result in the previous equations, we find 
{\it {two compatibility conditions}}

\be
\label{eq:cc1}
\cos 2\theta = \alpha \cos\theta 
\ee

\be
\label{eq:cc2}
F= \frac{4 \alpha}{1 -\alpha^2} E
\ee
 The  deduced matrix elements of Y, after imposing conditions (\ref{eq:dt}) are:
\beq
Y_{11}&=& \frac{\cos{2\theta}}{\alpha \sin\theta \sin{2\theta}} F\\
Y_{12}&=& \frac{1}{\alpha \sin\theta} \left( \frac{\cos{2\theta}}{
\sin{2\theta}}E -\frac{F}{2 \sin\theta} \right)\\
Y_{21}&=& \frac{F}{\sin\theta \sin{2\theta}} \\ 
Y_{22}&=& \frac{E - \cos\theta F}{\sin\theta \sin{2\theta}} \\
\eeq

Note that from experiment the ratio F/E has to be
\be
F/E =\lambda \approx .22
\ee 
The compatibility condition (\ref{eq:cc2}) led to a {\it{Cabbibo hierarchy}} for 
the vev ratio
\be
\alpha=\frac{v_2}{v_1} \approx \frac{\lambda}{4} \approx .055
\ee
Hence the condition (\ref{eq:cc1}) is satisfied for an angle
\be
\theta = \frac{\pi}{4} -\epsilon;
\ee
with
\be
\epsilon = \frac{\sqrt{2}}{4} \alpha + \frac{ \alpha^2}{8}+ {\cal{O}}(\alpha^3)
\ee  

Therefore, the matrices Y and I turn out to be, to order $\alpha^2$
\be
Y=\left(\begin{array}{ll} F &-3 E -\frac{3 \sqrt{2}}{8} F\\
\sqrt{2} F &\sqrt{2} E -\frac{7}{8} F \end{array} \right) \ \ \ \
I=\left(\begin{array}{ll} -F & 3 E +\frac{11 \sqrt{2}}{8} F\\
0 &\frac{3}{4} F \end{array} \right) 
\ee
Finally, we obtain, up to a multiplicative factor 
\be
Y=\left(\begin{array}{ll} \lambda &-3  -\frac{3 \sqrt{2}}{8} \lambda\\
\sqrt{2} \lambda &\sqrt{2}  -\frac{7}{8} \lambda \end{array} \right) \ \ \ \
I=\left(\begin{array}{ll} -\lambda & 3  +\frac{11 \sqrt{2}}{8} \lambda\\
0 &\frac{3}{4} \lambda \end{array} \right) 
\ee

The model analyzed serves as an existence proof of an alternative
 possible mechanism responsible for the structure of fermion mass matrices. 
 Instead of inducing a disparity between Yukawa couplings due to the 
 presence of Higgs multiplets with different couplings due to their different 
 representations, the possibility explored in this note indicates that the 
 phenomenologically required differentiation is achieved due to the necessary 
 rotation in the {\it{(Higgs isodoublet)- (left handed lepton doublet)}} 
 space induced 
 by $R$-parity violation. Independently of the features of the particular example 
 employed here it should be stressed that the circumstances suitable for this 
 mechanism to operate could most naturally arise in an effective $SU(3)\times 
 SU(2)\times U(1)$ theory resulting from Superstrings. Since higher Higgs 
 representations 
 are notoriously difficult to arise in Superstring embedable models 
 and since the equality of quark and lepton 
 Yukawas is required by string unification at the string scale without the 
 presence of any extra gauge symmetry, $R$-parity violation through the bilinear 
 Higgs-lepton couplings seems like an interesting possibility.
 
 {\bf{Aknowledgments}}
 
 The authors would like to aknowledge financial support from the TMR Network 
 ``Beyond the Standard Model".

\begin{thebibliography}{99}
\bibitem{XYZ}For a review see G. G. Ross {\it ``Grand Unified Theories"}
(Benjamin/Cummings, Menlo Park, California, 1985), see also \\
B. Ananthanarayan
 and P. Minkowski, hep-ph/9702279.

 \bibitem{XYW}M. Green, J. Schwarz and E. Witten, 
{\it ``Superstring Theory"}(Cambridge U. P. ,Cambridge 
1987).

 \bibitem{XYH}H. P. Nilles, {\it Phys. Rep. }
{\bf 110}(1984)1;\\ 
H. E. Haber and G. L. Kane, 
{\it Phys. Rep. }{\bf 117}(1985)75;\\
 A. B. Lahanas and D. V. Nanopoulos, {\it Phys. Rep.}
{\bf 145}(19887)1.

\bibitem{XYG}J. Ellis, S. Kelley and D. V. Nanopoulos,
 {\it Phys. Lett.}{\bf B260}(1991)131;\\
 U. Amaldi, W. De Boer and M. F\"{u}rstenau, 
{\it Phys. Lett. }{\bf B260}(1991)447;\\
 P. Langacker and M. Luo, {\it Phys. Rev. }{\bf D44}
(1991)817.

\bibitem{FR}For a review see C. Froggatt, Proceedings of the Fith Hellenic 
School and Workshops on Elementary Particles Physics, Corfu, 3 -24, September
1995. And references therein.
\bibitem{CH} M.S. Chanowitz, J. Ellis and M.K. Gaillard,\np{128}(1977) 506.

A. Buras, J. Ellis, M.K. Gaillard and D.V. Nanopoulos, \np{135}(1978) 66.
   
\bibitem{GJ} H.Georgi and C.Jarlskog, \plett{86}(1979) 297.
\bibitem{RA} P. Ramond, R.G. Roberts and G.G. Ross, \np{406}(1993) 19.
\bibitem{AN}G.~Anderson,
S.~Raby, S.~Dimopoulos, L.~J.~Hall, and G.~D.~Starkman, \prd{49}(1994) 3660.

K.~S. Babu and R.~N. Mohapatra, \prl{74}(1995) 2418.

\bibitem{KT} K. Tamvakis, \plett{383}(1996)307.
\eb  
\ed